\documentclass[10 pt, journal]{IEEEtran}

\makeatletter
\def\ps@headings{%
	\def\@oddhead{\mbox{}\scriptsize\rightmark \hfil \thepage}%
	
	\def\@evenhead{\scriptsize\thepage \hfil \leftmark\mbox{}}%
	
	\def\@oddfoot{}%
	
	\def\@evenfoot{}}
	
\makeatother
\usepackage{cite}
\usepackage{gensymb}
\usepackage{float}  
\usepackage{epsfig,bm,amsmath,amssymb,graphicx,setspace}
\usepackage[numbers,sort&compress]{natbib}
\usepackage{color,placeins}
\usepackage{multirow}
\usepackage{algorithm}
\usepackage{algpseudocode}
\usepackage{colortbl}
\usepackage{color}
\usepackage{dblfloatfix}
\usepackage{turnstile}
\usepackage{multicol}
\usepackage{array}

\usepackage{enumerate}
\usepackage{turnstile}
\usepackage{subcaption}
\usepackage{arydshln,paralist}
\usepackage{soul,tabularx,booktabs}
\usepackage{csquotes}
\usepackage{multirow}
\usepackage{adjustbox}
\usepackage{float}
\usepackage{color, colortbl}
\definecolor{LightCyan}{rgb}{0.88,1,1}
\usepackage{xcolor}
\usepackage{verbatim}
\usepackage[colorinlistoftodos]{todonotes} 
\usepackage{rotating}
\definecolor{usethiscolorhere}{rgb}{0.86666,0.78431,0.78431}
\usepackage{tikz}
\usetikzlibrary{mindmap}
\usetikzlibrary{calc,positioning}
\usepackage{lipsum,adjustbox}
\usetikzlibrary{decorations.pathmorphing}

\usepackage{pifont}

\makeatother

\usepackage{amsmath}
\usepackage[hyphens]{url}

\begin{document}

\title{An Analysis of Energy Consumption and Carbon Footprints of Cryptocurrencies and Possible Solutions}

\author{Varun Kohli$^a$, Sombuddha Chakravarty$^b$, Vinay Chamola$^{*,b}$, Kuldip Singh Sangwan$^c$, Sherali Zeadally$^d$

\thanks{$^a$ Department of Electrical and Computer Engineering Engineering, National University of Singapore, Singapore (email: varun.kohli@u.nus.edu)}%
\thanks{$^b$ Department of Electrical and Electronics Engineering \& APPCAIR, BITS-Pilani, Pilani Campus, 333031, India  (email: f2016165p@alumni.bits-pilani.ac.in, vinay.chamola@pilani.bits-pilani.ac.in)}%
\thanks{$^c$ Department of Mechanical Engineering, BITS-Pilani, Pilani Campus, 333031, India (email: kss@pilani.bits-pilani.ac.in)}%
\thanks{$^d$ College of Communication and Information, University of Kentucky, Lexington, KY 40506-0224 (email: szeadally@uky.edu).}}%

\maketitle{}

\begin{abstract}
There is an urgent need to control global warming caused by humans to achieve a sustainable future. $CO_2$ levels are rising steadily and while countries worldwide are actively moving toward the sustainability goals proposed during the Paris Agreement in 2015, we are still a long way to go from achieving a sustainable mode of global operation. The increased popularity of cryptocurrencies since the introduction of Bitcoin in 2009 has been accompanied by an increasing trend in greenhouse gas emissions and high electrical energy consumption. Popular energy tracking studies (e.g., Digiconomist and the Cambridge Bitcoin Energy Consumption Index (CBECI)) have estimated energy consumption ranges of 29.96 TWh to 135.12 TWh and 26.41 TWh to 176.98 TWh respectively for Bitcoin as of July 2021, which are equivalent to the energy consumption of countries such as Sweden and Thailand. The latest estimate by Digiconomist on carbon footprints shows a 64.18 Mt$CO_2$ emission by Bitcoin as of July 2021, close to the emissions by Greece and Oman. This review compiles estimates made by various studies from 2018 to 2021. We compare with the energy consumption and carbon footprints of these cryptocurrencies with countries around the world, and centralized transaction methods such as Visa. We identify the problems associated with cryptocurrencies, and propose solutions that can help reduce their energy usage and carbon footprints. Finally, we present case studies on cryptocurrency networks namely, Ethereum 2.0 and Pi Network, with a discussion on how they solve some of the challenges we have identified.
\end{abstract}


\section{Introduction}
\label{sec:introduction}

The past century has witnessed a steady rise in atmospheric Green House Gas (GHG) levels with nearly 584 Gt $CO_2$ from fossil fuels, land use change and industrial activity contributing to 0.9\degree C of global temperature increase since 1960 \cite{mora2018bitcoin}. $CO_2$ levels have increased from 250ppm in 1960 to 400ppm in 2020 and current average trends show a rise in natural disasters caused by high temperatures and droughts \cite{zandalinas2021global}. Day and night temperatures have increased worldwide, and the average global temperatures are expected to go up by 3-5\degree C by 2100 according to the Intergovernmental Panel on Climate Change (IPCC) \cite{IPCC}. Evidence suggests a change in the lengths of seasons across the globe due to global warming. Summers in the mid-high latitudes have lengthened while the winters have shortened, as well as shorter spring and autumn periods \cite{wang2021changing}. It has been predicted that even if the GHG levels do not increase beyond current levels, summers will last for nearly half a year while winters will be less than two months long by 2100. 

In 2015, leaders from 197 countries settled upon the Paris Agreement with the aim to keep global warming caused by human beings under 2\degree C \cite{rogelj2016paris}. This is already a difficult task given the increase in population, energy consumption and the lack of environment friendly policies by governments worldwide. USA, China, Japan, Germany, and India which have been the main ecological footprint hotspots since 2019, also correspond to the top GHG emission nations across the world \cite{sarkodie2021environmental}. Since 2009, various cryptocurrencies have emerged starting with Bitcoin which was the first well-known application of Satoshi Nakamoto's blockchain technology introduced in 2008 \cite{nakamoto2008bitcoin}. It soon became the biggest cryptocurrency in the world with a market capitalization of USD \$ 614.9 billion as of July 2021 among the 5,655 known cryptocurrencies \cite{coin2021crypto}: Ethereum, Tether, Binance, Cardano, and Dogecoin to name a few; and together they account for to a total market capitalization of USD \$1.39 trillion. Millions of transactions are made every single day to exchange these currencies and their stock markets and operations run 24/7 \cite{stoll2019carbon}. The electrical energy consumption of cryptocurrencies is over-proportionate compared to their technical performance \cite{sedlmeir2020energy} and despite their promising applications, cryptocurrencies have also been contributors responsible for global warming due to their high carbon footprint \cite{de2021bitcoin}. It has been predicted that Bitcoin alone can raise the global temperatures by 2\degree C within the next three decades \cite{mora2018bitcoin}.

\begin{figure*}[!t]
        \vspace{-0.1in}
        \centering
        \includegraphics[width = \textwidth]{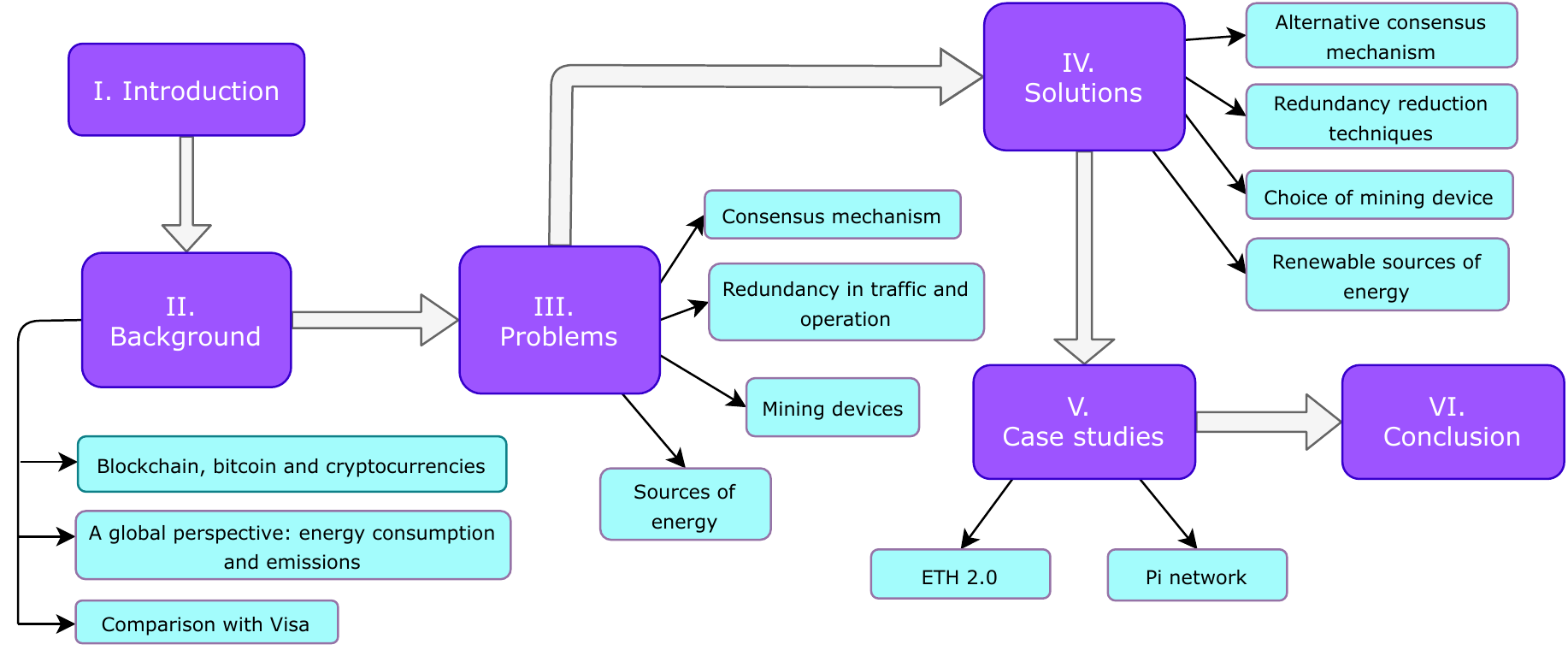}
        \caption{The flow of this review.}
        \label{flow}
        \vspace{-0.1in}
\end{figure*}

Due to the distributed nature of cryptocurrency networks, obtaining close estimates of electrical energy consumption and carbon footprints is a difficult task. The main source of uncertainty is the mining equipment used \cite{houy2019rational} and the source of energy \cite{koomey2019estimating}. The minimum and maximum power demand estimates for the Bitcoin network according to various studies conducted between 2014 and 2018 have been compiled in \cite{kufeouglu2019bitcoin}. Estimates in the range of 2.5 GW to 7.67 GW \cite{de2018bitcoin}, 1.3 GW to 14.8 GW and 15.47 TWh to 50.24 TWh \cite{kufeoglu2019energy}, and 22 TWh to 105 TWh \cite{zade2019bitcoin} were made for Bitcoin in 2018. The power consumption was later estimated to be 4.3 GW in March 2020, nearly a 68\% share of the top 20 cryptocurrencies drawing a total of 6.5 GW \cite{gallersdorfer2020energy}. This was done without considering auxiliary losses caused by the cooling and mining equipment, and with that premise, the true power draw is expected to be higher. With the consideration that only 20 cryptocurrencies were used in this study, the actual cryptocurrency network power consumption of the 5,654 cryptocurrencies \cite{coin2021crypto} would be much higher than their estimate. Among the latest data on consumption, the University of Cambridge Bitcoin Energy Consumption Index (CBECI) \cite{uoc2021bitcoin} shows theoretical maximum and minimum power consumptions of 26.09 TWh to 174.82 TWh respectively, with an estimate of 69.63 TWh. A study from early 2021 \cite{sedlmeir2020energy} showed a range of 60 TWh to 125 TWh per year for Bitcoin, 15 TWh for Ethereum and 100 TWh for Bitcoin Cash. A sensitivity-based method used by Alex de Vries in early 2021 \cite{de2021bitcoin} factored into the Bitcoin market cost, electricity cost and the percentage of miners' income spent on electricity. The results showed the Bitcoin network energy consumption to be up to 184 TWh. His famous blog, Digiconomist founded in 2014 \cite{digi} estimates the energy consumption of Bitcoin and Ethereum to be 135.12 TWh and 55.01 TWh respectively as of July 2021.  

As a consequence of high electrical energy consumption, cryptocurrencies have also been found to have high carbon footprints. The carbon footprint of Bitcoin alone was estimated to be 63 Mt$CO_2$ in 2018 \cite{kohler2019life} and 55 Mt$CO_2$ in 2019 \cite{stoll2019carbon}. Another study in 2018 \cite{sriman2021blockchain} stated a footprint of 38.73 Mt$CO_2$ which was equivalent to Denmark, over 700,000 Visa transaction and nearly 49,000 hours of YouTube viewing. Alex de Vries showed the consumption to be up to 90.2 Mt$CO_2$ \cite{de2021bitcoin} early in 2021 with an estimate of 64.18 Mt$CO_2$ \cite{digi2021bitcoin}. Along similar lines, Digiconomist also calculated a 26.13 Mt$CO_2$ footprint for Ethereum in July 2021. The 3rd Global Cryptoasset Benchmarking Study (GCBS) conducted by the University of Cambridge in 2020 \cite{blandin20203rd} found an average of 39\% of renewable energy share in Proof of Work (PoW) mining while a contesting result was found in a 2018 study \cite{bendiksen2018bitcoin} with a 78\% share of renewable energy. But considering the high carbon footprints for these cryptocurrencies, we can infer that there is still a considerable load on non-renewable sources of energy such as fossil fuels. 

From the discussion so far, we found that the energy consumption and carbon footprint of cryptocurrencies are very high. We show later in this work that these metrics are close to if not more than those of several countries and much of this high energy consumption stems from mechanisms used by many of the cryptocurrency implementations. Figure-\ref{flow} presents the organization of this review paper. We summarize the main research contributions of this review as follows:
\begin{itemize}
\item{We present a global perspective on energy consumption and carbon footprints by the two most popular cryptocurrencies namely, Bitcoin and Ethereum. We also present a comparison of energy consumption and carbon emissions of Bitcoin, Ethereum, and the card payment system Visa.}
\item{We identify four underlying factors responsible for high energy consumption and carbon emissions of Bitcoin and Ethereum.}
\item{We discuss possible solutions to address the factors that result in high energy consumption and carbon emissions for cryptocurrencies such as Bitcoin and Ethereum. Additionally, we discuss two case studies on work-in-progress solutions.}
\end{itemize}

\begin{figure*}[!t]
        \centering
        \includegraphics[width = \textwidth]{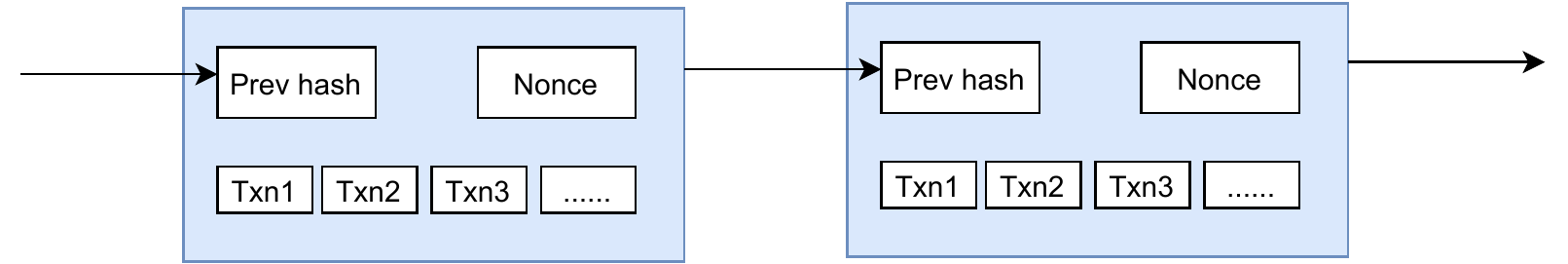}
        \caption{A block diagram depicting the structure of a blockchain.}
        \label{blockchain}
        \vspace{-0.1in}
\end{figure*}

\begin{table*}[t]
\centering
\caption{Ranking Bitcoin and Ethereum among countries based on annual electrical energy consumption as of July 2021 \cite{digi2021bitcoin,world2021pop,digi2021ethereum,eia2019energy,wiki2019energy} (Note: N.A. stands for Not Available).}
\resizebox{1.5\columnwidth}{!}{
\begin{tabular}{|l|l|r|r|r|}
\hline
\multicolumn{1}{|c|}{\textbf{Rank}} & \multicolumn{1}{c|}{\textbf{Country}} & \multicolumn{1}{c|}{\textbf{\begin{tabular}[c]{@{}c@{}}Population \\ (Millions) \cite{world2021pop}\end{tabular}}} & \multicolumn{1}{c|}{\textbf{\begin{tabular}[c]{@{}c@{}}Energy (TWh) \\  \cite{digi2021bitcoin,digi2021ethereum,eia2019energy,wiki2019energy}\end{tabular}}} & \multicolumn{1}{c|}{\textbf{\begin{tabular}[c]{@{}c@{}}Share \\ (\%)\end{tabular}}} \\ \hline
0                                   & World                                 & 7,878.2                                                                                         & 23,398.00                                                                             & 100.00                                                                              \\ \hline
1                                   & China                                 & 1,444.9                                                                                         & 7,500.00                                                                              & 32.05                                                                               \\ \hline
2                                   & U.S.A                                 & 332.9                                                                                          & 3,989.60                                                                              & 17.05                                                                               \\ \hline
3                                   & India                                & 1,366.4                                                                                          & 1,547.00                                                                              & 6.61                                                                               \\ \hline

20                                  & Taiwan                                & 23.8                                                                                           & 237.55                                                                                & 1.01                                                                                \\ \hline
21                                  & Vietnam                               & 98.2                                                                                           & 216.99                                                                                & 0.92                                                                                \\ \hline
22                                  & South Africa                          & 60.1                                                                                           & 210.30                                                                                & 0.89                                                                                \\ \hline \rowcolor{LightCyan}
23                                  & Bitcoin + Ethereum                    & N.A.                                                                                           & 190.13                                                                                & 0.81                                                                                \\ \hline
24                                  & Thailand                              & 69.9                                                                                           & 185.85                                                                                & 0.79                                                                                \\ \hline
25                                  & Poland                                & 37.80                                                                                          & 153.00                                                                                & 0.65                                                                                \\ \hline
26                                  & Egypt                                 & 104.3                                                                                          & 150.57                                                                                & 0.64                                                                                \\ \hline
27                                  & Malaysia                              & 3.1                                                                                            & 147.21                                                                                & 0.62                                                                                \\ \hline \rowcolor{LightCyan}
28                                  & Bitcoin                               & N.A.                                                                                           & 135.12                                                                                & 0.57                                                                                \\ \hline
29                                  & Sweden                                & 10.2                                                                                           & 131.79                                                                                & 0.56                                                                                \\ \hline

49                                  & Switzerland                           & 8.7                                                                                            & 56.35                                                                                 & 0.24                                                                                \\ \hline \rowcolor{LightCyan}
50                                  & Ethereum                              & N.A.                                                                                         & 55.01                                                                                 & 0.24                                                                                \\ \hline
51                                  & Romania                               & 19.1                                                                                           & 55.00                                                                                 & 0.23                                                                                \\ \hline
\end{tabular}}
\label{energy-comp}
\end{table*}

\begin{table*}[t]
\centering
\caption{Ranking of Bitcoin and Ethereum among countries based on annual carbon footprint as of July 2021 \cite{digi2021bitcoin,digi2021ethereum,iea2018carbon,world2021pop}.}
\resizebox{1.5\columnwidth}{!}{
\begin{tabular}{|l|l|r|r|r|}
\hline
\multicolumn{1}{|c|}{\textbf{Rank}} & \multicolumn{1}{c|}{\textbf{Country}} & \multicolumn{1}{c|}{\textbf{\begin{tabular}[c]{@{}c@{}}Population \\ (Millions) \cite{world2021pop} \end{tabular}}} & \multicolumn{1}{c|}{\textbf{\begin{tabular}[c]{@{}c@{}}Emission (Mt$CO_2$) \\  
\end{tabular}}} & \multicolumn{1}{c|}{\textbf{\begin{tabular}[c]{@{}c@{}}Share \\ (\%)\end{tabular}}} \\ \hline
0                                            & World                                 & 7,878.2                                                                                         & 37,077.40                                                                                 & 100.00                                                                              \\ \hline
1                                            & China                                 & 1,444.9                                                                                         & 10,060.00                                                                                 & 27.13                                                                               \\ \hline
2                                            & U.S.A                                 & 332.9                                                                                          & 5410.00                                                                                   & 14.59                                                                               \\ \hline
3                                            & India                                 & 1,336.4                                                                                          & 2,300.00                                                                                   & 6.2                                                                               \\ \hline

38                                           & Nigeria                               & 211.3                                                                                          & 104.30                                                                                    & 0.28                                                                                \\ \hline
39                                           & Czech Republic                        & 10.7                                                                                           & 100.80                                                                                    & 0.27                                                                                \\ \hline
40                                           & Belgium                               & 11.6                                                                                           & 91.20                                                                                     & 0.24                                                                                \\ \hline \rowcolor{LightCyan}
41                                           & Bitcoin + Ethereum                    & N.A.                                                                                           & 90.31                                                                                     & 0.24                                                                                \\ \hline
42                                           & Kuwait                                & 4.3                                                                                            & 87.80                                                                                     & 0.23                                                                                \\ \hline
43                                           & Qatar                                 & 2.9                                                                                            & 87.00                                                                                     & 0.23                                                                                \\ \hline
49                                           & Oman                                  & 5.2                                                                                            & 68.80                                                                                     & 0.18                                                                                \\ \hline \rowcolor{LightCyan}
50                                           & Bitcoin                               & N.A.                                                                                           & 64.18                                                                                     & 0.17                                                                                \\ \hline
51                                           & Greece                                & 10.3                                                                                           & 61.60                                                                                     & 0.16                                                                                \\ \hline
76                                           & Tunisia                               & 11.94                                                                                          & 26.20                                                                                     & 0.07                                                                                \\ \hline \rowcolor{LightCyan}
77                                           & Ethereum                              & N.A.                                                                                           & 26.13                                                                                     & 0.07                                                                                \\ \hline
78                                           & SAR                                   & 17.9                                                                                           & 25.80                                                                                     & 0.06                                                                                \\ \hline
\end{tabular}}
\label{$CO_2$-comp}
\end{table*}

\section{Background}
\label{sec:background}
This section presents a brief overview of blockchain technology and cryptocurrencies. It provides a global perspective of energy consumptions and carbon emissions of the two biggest cryptocurrencies namely, Bitcoin and Ethereum, and compares them to the centralized banking system, Visa.  

\subsection{Blockchain, Bitcoin and Cryptocurrencies}

Blockchain is a disruptive technology of distributed ledgers which was created by Satoshi Nakamoto in 2008 \cite{nakamoto2008bitcoin}. A blockchain is a database that chronologically stores information in "blocks". These blocks have a storage capacity of information that consists of the stored information, a time-stamp, the hash value of the previous block, and a unique identification number called the nonce. Once a block has been filled, it is added or "chained" onto the previously filled block thereby creating a "blockchain" as Figure-\ref{blockchain} shows. In addition, any changes to a block are detected by the hash value for that block making it easy to identify fraud \cite{nofer2017blockchain}. Blockchain offers many benefits. First, it stores data chronologically and securely, with a copy of the ledger stored on every node in the cryptocurrency network. Second, the functionality of the network is maintained even if a few participating nodes are removed or malfunction. Third, peer-peer trust is maintained through the consensus mechanism, which removes the need for intermediaries that may not be trustworthy. Blockchain finds applications in various areas such as logistics and supply chain \cite{zeadally2019blockchain, hassija2020survey}, e-commerce \cite{ometov2020overview}, education \cite{grather2018blockchain}, healthcare \cite{ismail2019lightweight}, governance \cite{olnes2017blockchain} and others \cite{pilkington2016blockchain}. It can also be used in  telecommunication technology \cite{praveen2020blockchain}, stock exchange \cite{bansal2019smart}, industrial IoT \cite{alladi2019blockchain}, smart city development \cite{hassija2020traffic, hassija2020parking}, energy management \cite{miglani2020blockchain}, Unmanned Aerial Vehicles (UAV) \cite{alladi2020applications}, and smart grids \cite{alladi2019blockchain2}. But the most successful application has been in the banking sector \cite{hileman2017global} with the rise of over 5,000 cryptocurrencie as of July 2021 \cite{coin2021crypto}. 

Bitcoin, as described by Satoshi Nakamura, is a peer-peer electronic cash system in which the double spending prevention process is decentralized across various nodes through a consensus protocol. All Bitcoin transactions are time-stamped, and any double spending attempts are rejected. "Bitcoin Miners" play a major role in maintaining consensus over the ledger's state through the PoW (discussed in depth in Section-\ref{sec:problems}) in which they compete with others on the cryptocurrency network to solve resource intensive cryptographic problems to earn the right to add their proposed block onto the chain. The difficulty of the puzzle changes over time to maintain the time to mine a block at nearly 10 minutes \cite{antonopoulos2014mastering}. The miners invest in higher computational power in order to not be left behind in the race of pushing their blocks onto the ledger. Successful attempts are awarded a certain quantity of Bitcoin (BTC) as a reward for each block solved. The reward is halved after every 210,000 blocks, in order to maintain a steady synthetic inflation until the 21 million possible BTC is in circulation \cite{berg2020proof,mora2018bitcoin}. The reward per block has been 6.25 BTC since the most recent halving that occurred on May 11, 2020 \cite{bitcoin2021clock}. With nearly 140,000 blocks left to mine, the next halving is expected to occur on March 26, 2024.

Another popular blockchain network, Ethereum, introduced the concept of a programmable network. Ethereum supports the cryptocurrency Ether (ETH) which has the second highest market capitalization \cite{coin2021crypto}. With the development of the Ethereum Virtual Machine (EVM), the concept of smart contracts (i.e. the automatic execution of contracts when certain conditions are met) was proposed. However, as is the case in Bitcoin, Ethereum is also based on the PoW consensus algorithm and therefore it is associated with the same issues of electrical energy consumption and carbon footprints. Ethereum has proposed Ethereum 2.0 in order to address most of the issues with BTC and ETH which we discuss in more detail in Section-\ref{sec:case}.

\subsection{A Global Perspective: Energy Consumption and $CO_2$ Emissions}
Table-\ref{energy-comp} shows the comparison of electrical energy consumption of Bitcoin and Ethereum obtained from Digiconomist \cite{digi2021bitcoin,digi2021ethereum}. We obtained the country-wise consumption and population data from the U.S. Energy Information Administration database \cite{eia2019energy} and Worldometer \cite{world2021pop} respectively. We calculated the percentage Share of energy consumption as follows:

\begin{equation}
    Share = \frac{Energy_i}{Energy_{w}}\times100
\end{equation}

where $Energy_i$ is the energy consumption of the country at rank $i$ and $Energy_w$ corresponds to the total energy consumption of the world as the table shows. Accordingly, the estimates of the total electrical energy consumption share of Bitcoin and Ethereum are 0.58\% and 0.23\%. They rank 28th and 50th with 135.12 TWh and 55.01 TWh of consumption respectively. The University of Cambridge has also arrived at a close estimate of 0.6\% for Bitcoin \cite{uoc2021bitcoin} which supports these calculations. The consumption by Bitcoin is comparable to Sweden (131.79 TWh, 0.56\%), while that by Ethereum is nearly the same as Romania (55 TWh, 0.23\%). Considering the high rated power share of 79.85\% for these two cryptocurrencies among all in circulation as of March 2020 \cite{gallersdorfer2020energy}, the data for the two cryptocurrencies as a single entity has also been considered to obtain a holistic representation. It is worth noting that they together rank 23rd in the world and consume a total of 190.13 TWh of energy annually with a share of 0.81\%, which is equivalent to Thailand (185.85 TWh, 0.79\%). 

Table-\ref{$CO_2$-comp} presents a similar ranking, but this time based on the annual $CO_2$ emissions. Data on the emissions of various countries was obtained from the International Energy Agency database \cite{iea2018carbon}. The percentage Share has been calculated in the same manner as for energy consumption. It can be observed from the table that Bitcoin ranks 50th in emissions among the 143 countries in this database, with 64.18 Mt$CO_2$ of emissions and a share of 0.17\%. These values are close to those of Oman (68.8 Mt$CO_2$, 0.18\%) and Greece (61.6 Mt$CO_2$, 0.16\%). The statistics for Ethereum are also significant, with a rank of 77, emissions of 26.13 Mt$CO_2$ and a 0.07\% global share, which is comparable to Tunisia (26.2 Mt$CO_2$, 0.07\%). When the two cryptocurrencies are considered together, they rank 41 in the world with emissions of 90.31 Mt$CO_2$ and a global share of 0.24\% which is nearly the same as Belgium (91.2 Mt$CO_2$, 0.24\%).

\begin{table}[t]
\centering
\caption{Energy consumption and carbon footprints of Bitcoin, Ethereum and Visa (total) as of July 2021 \cite{digi2021bitcoin,digi2021ethereum,impakter2018Visa}.}
\resizebox{\columnwidth}{!}{
\begin{tabular}{|l|r|r|r|c|}
\hline
\multicolumn{1}{|c|}{\textbf{\begin{tabular}[c]{@{}c@{}}Transaction \\ method\end{tabular}}} & \multicolumn{1}{c|}{\textbf{\begin{tabular}[c]{@{}c@{}}Market cap \\ (\$ Billion)\end{tabular}}} & \multicolumn{1}{c|}{\textbf{\begin{tabular}[c]{@{}c@{}}Transactions/day \\ (Million)\end{tabular}}} & \multicolumn{1}{c|}{\textbf{\begin{tabular}[c]{@{}c@{}} Emission \\ (Mt$CO_2$)\end{tabular}}} & \multicolumn{1}{c|}{\textbf{\begin{tabular}[c]{@{}c@{}}Energy consumption \\ (TWh)\end{tabular}}} \\ \hline
Bitcoin \cite{digi2021bitcoin}                                                                                    & 617.05                                                                                           & 0.4                                                                                                 & 64.18                                                                                              & 135.12                                                                               \\ \hline
Ethereum \cite{digi2021ethereum}                                                                                 & 247.8                                                                                            & 1.23                                                                                                & 26.13                                                                                              & 55.01                                                                                \\ \hline
Visa \cite{impakter2018Visa}                                                                                       & 520.62                                                                                           & 500                                                                                                 & 62,400                                                                                             & 197.57                                                                               \\ \hline
\end{tabular}}
\label{comp-total}
\end{table}

\begin{table}[t]
\centering
\caption{Comparison of energy consumption and carbon footprints per transaction for Bitcoin, Ethereum and Visa as of July 2021 \cite{digi2021bitcoin,digi2021ethereum}.}
\resizebox{0.8\columnwidth}{!}{
\begin{tabular}{|l|r|c|}
\hline
\multicolumn{1}{|c|}{\textbf{\begin{tabular}[c]{@{}c@{}}Transaction \\ method\end{tabular}}} & \multicolumn{1}{c|}{\textbf{\begin{tabular}[c]{@{}c@{}}Emission \\ (Kg$CO_2$)\end{tabular}}} & \multicolumn{1}{c|}{\textbf{\begin{tabular}[c]{@{}c@{}}Energy consumption \\ (kWh)\end{tabular}}} \\ \hline
Bitcoin \cite{digi2021bitcoin}                                                                                    & 844.13                                                                                      & 1777.11                                                                                          \\ \hline
Ethereum \cite{digi2021ethereum}                                                                                    & 59.55                                                                                       & 125.36                                                                                           \\ \hline
Visa \cite{digi2021ethereum}                                                                                       & 0.00045                                                                                     & 0.0015                                                                                           \\ \hline
\end{tabular}}
\label{comp-trans}
\end{table}

\subsection{Comparison with Visa}

Table-\ref{comp-total} and Table-\ref{comp-trans} present the data available on the energy consumption and $CO_2$ emissions of Bitcoin \cite{digi2021bitcoin}, Ethereum \cite{digi2021ethereum} and Visa \cite{impakter2018Visa,digi2021ethereum}. 

\begin{figure}[!t]
        \vspace{-0.1in}
        \centering
        \includegraphics[width = \columnwidth]{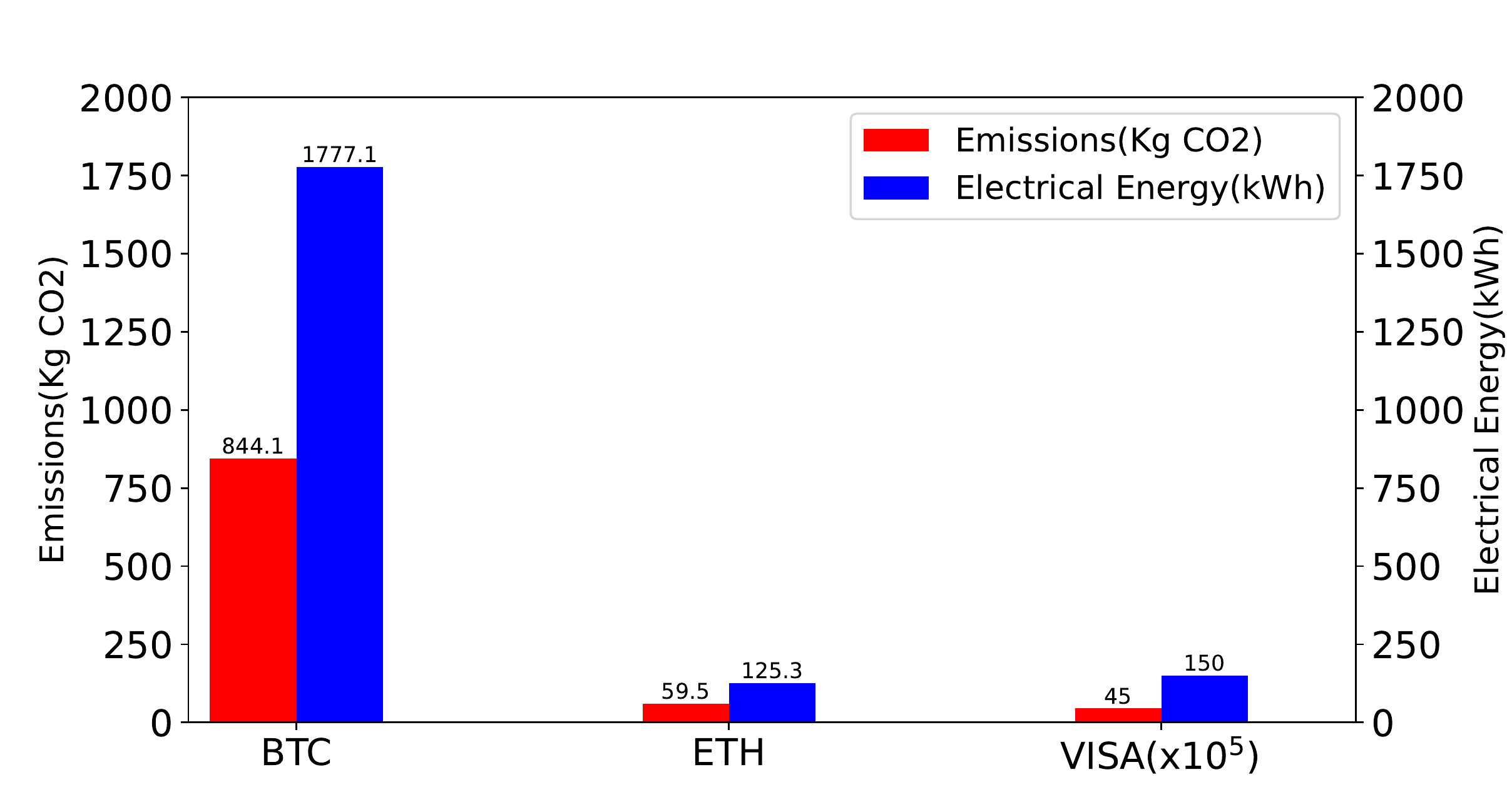}
        \caption{Electrical energy consumption and $CO_2$ emissions per transaction for Bitcoin, Ethereum and Visa \cite{digi2021bitcoin,digi2021ethereum}.}
        \label{btc-eth-Visa}
        \vspace{-0.1in}
\end{figure}

Table-\ref{comp-total} shows the annual energy consumption and emission values for the three transactions methods,considering all sources of consumption in Visa. While at first glance it may seem that the total $CO_2$ emission and energy consumption are comparatively high for Visa, it is worth pointing out that the number of daily transactions occurring in the Bitcoin and Ethereum networks is 0.4 million and 1.25 million, i.e. 0.08\% and 0.25\% respectively of the 500 million daily Visa transactions. This implies the over-proportionate consumption in cryptocurrencies which are relatively nascent transaction methods. In addition, the total metrics for Visa have been calculated considering all requirements to run the cooperation offices such as office and server electricity, and commute. Table-\ref{comp-trans} shows the per-transaction estimates for the three transaction methods considering only the computational costs. From the table we observe that the energy consumption and $CO_2$ emission per transaction are very high for Bitcoin and Ethereum. Figure-\ref{btc-eth-Visa} presents a visual comparison between these metrics per transaction. Energy consumption and $CO_2$ emissions for Visa have been plotted by raising their values by a factor of $10^5$. Accordingly, Table-\ref{BE} shows the Break Even ($BE$) values that correspond to the number of Visa transactions that can occur to have total energy consumption and $CO_2$ emission equal to a single transaction of these cryptocurrencies. We calculate $BE$ as follows: 

\begin{equation}
    \label{eq:BE}
    BE_{Visa/i}^M = \frac{M_i}{M_{Visa}} 
\end{equation}

\begin{table}[b]
\centering
\caption{Break Even (BE) count for the number of Visa transactions per Bitcoin and Ethereum transaction as of July 2021, obtained from Equation-\ref{eq:BE}.}
\resizebox{0.95\columnwidth}{!}{
\begin{tabular}{|l|l|l|}
\hline
\textbf{Category}   & \textbf{$BE^{Energy consumption}$} & \textbf{$BE^{CO_2 emission}$} \\ \hline
Visa/Bitcoin   & 1,195,657         & 1,870,875         \\ \hline
Visa/Ethereum & 83,574           & 132,334         \\ \hline
\end{tabular}}
\label{BE}
\end{table}

where $BE_{Visa/i}^M$ is the $BE$ value for Visa with cryptocurrency $i$, which is either Bitcoin or Ethereum. $M$ corresponds to the metric in consideration, energy consumption or $CO_2$ emissions. As Table-\ref{BE} shows, it takes 1,195,657 Visa transactions to use the same amount of electrical energy as one transaction of Bitcoin. Similarly, it takes 83,574 Visa transactions to generate the same carbon footprint as a single transaction of Bitcoin. Similarly the $BE$ counts of Visa to Ethereum are 83,574 for energy consumption and 132,334 for carbon footprint.   

\begin{figure}[b]
        \vspace{-0.1in}
        \centering
        \includegraphics[width = \columnwidth]{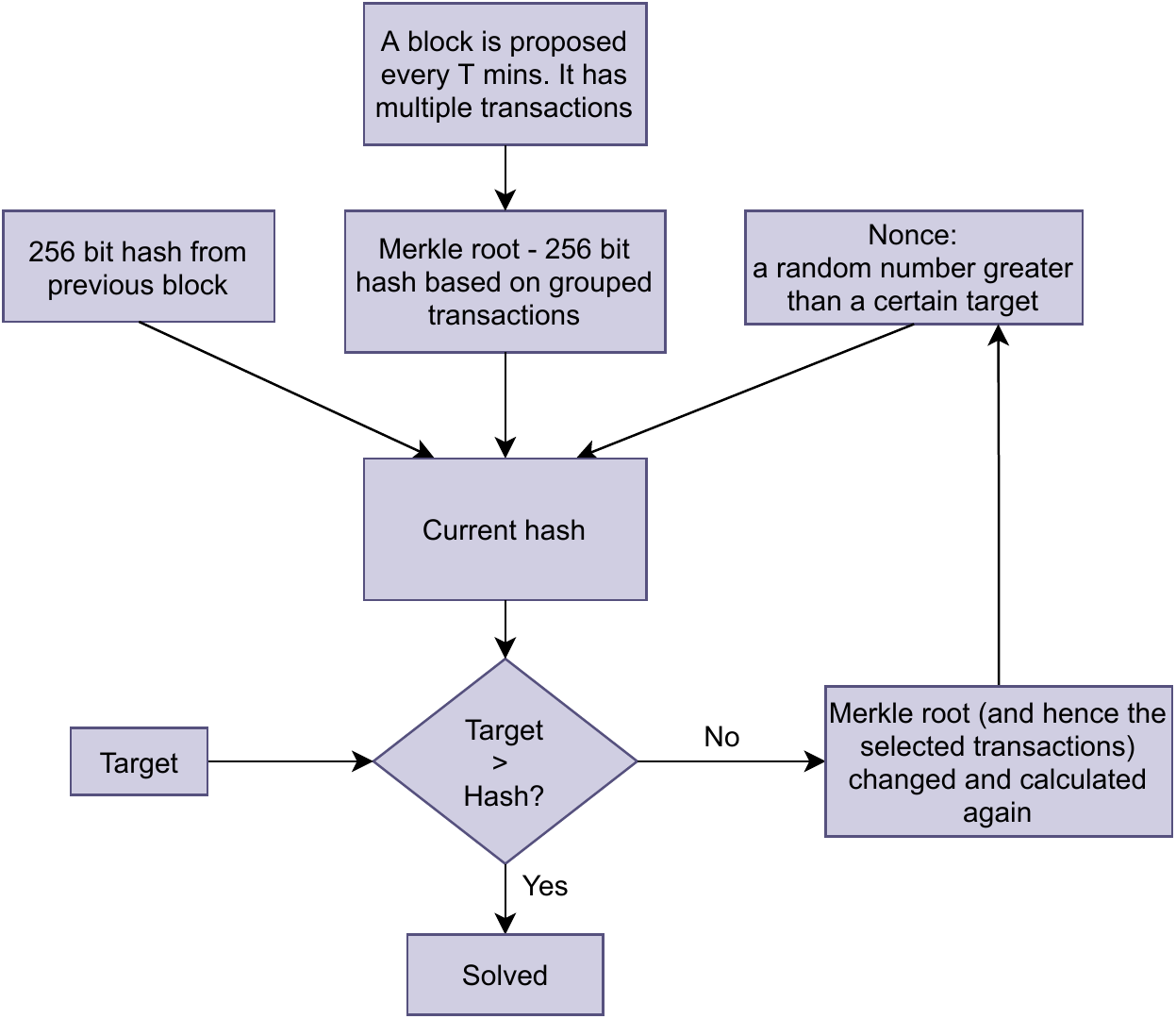}
        \caption{The mining process in Bitcoin.}
        \label{mining-flow}
        \vspace{-0.1in}
\end{figure}

\section{Problems}
\label{sec:problems}
Based on a review of past studies, we have identified four major responsible for the high energy consumption and $CO_2$ emissions in cryptocurrencies, namely: the Proof of Work consensus mechanism, redundancy in operation and traffic, mining devices, and the energy sources. This section discusses these issues so that future development in cryptocurrencies can take them into consideration.

\subsection{Consensus Mechanism: Proof of Work}

PoW was the first consensus mechanism proposed for blockchain networks \cite{nakamoto2008bitcoin}. Paul Haunter, a contributor of Ethereum, acknowledged the high energy requirements of PoW \cite{ieee2019ethereum} being the reason for the development of Ethereum 2.0 which we discuss in more detail in Section-\ref{sec:case}. While redundancy in the operation and traffic of cryptocurrency networks is also a contributor to energy consumption (as we discuss in the next subsection), the transactions themselves do not consume as much energy as the PoW process does. It has been proven that PoW mining has high computational needs and thus imposes major limitations on the continuous use and scalability of cryptocurrencies \cite{mishra2017energy,hassijaframework}. Recent research estimates that PoW mining in Bitcoin consumes nearly 18GW of power for 100 million transactions a week \cite{mishra2017energy} making the practical use Bitcoin questionable. Based on current trends, a study from 2021 has predicted that, because of the rapid growth of cryptocurrencies, PoW mining processes in China alone will consume nearly 300 TWh of electrical energy and generate 130 Mt$CO_2$ by 2024 \cite{jiang2021policy}. To understand why it is an important energy issue, we need to first understand its operation.

Figure-\ref{mining-flow} shows the mining process in Bitcoin using the PoW. Each new block proposed every T minutes is given a hash that is computed using the 256-bit hash of the previous block, the Nonce and the Merkle root using the equation:

\begin{equation}
SHA256(H_{prev} + M_{B} + Nonce) \leq Target
\end{equation}

where SHA256 is the hash function, $H_{prev}$ refers to the 256-bit hash of previous block, Nonce is a one time use positive number and $M_{B}$ is the Merkle root. Once the hash has been calculated, it is compared with the target hash value. This target value is set to increase the difficulty of mining so as to maintain a constant time for the block to be added to the chain. This time is set to 10 minutes for Bitcoin. If the hash is higher than the target, the Merkel root is changed, the nonce is re-calculated and another hash is generated. This process is repeated until the miner reaches a hash value below the target value set.

It is computationally expensive to find the nonce and therefore provides the proof of the amount of computational power put in by the miner, thereby giving this consensus mechanism the name PoW. Since the solution searching process cannot be sped up by parallelization and alternative algorithms \cite{wang2019survey}, a miner's share of reward can be equated to the share of computational power owned in the cryptocurrency network \cite{de2021bitcoin}. As mining becomes harder over time, the PoW becomes an arms race of computational power and resources because miners with more powerful devices compute more hashes per second.

\subsection{Redundancy in Traffic and Operation}

While PoW blockchains have energy problems which stem mainly from the consensus mechanism, energy consumption due to redundant operations and network traffic becomes more relevant in non-PoW blockchains. It arises from the system storing the complete ledger on all nodes in the network \cite{jia2018elasticchain}. In addition, each node performs operations associated with the transactions independently, based on the available transaction information. Additionally, redundant network traffic is another contributor to this problem \cite{zhang2021traffic}. Redundancy reduces the efficacy of the system \cite{zhang2021traffic} while also increasing the total electrical energy consumption \cite{sedlmeir2020energy}. 

As stated in \cite{sedlmeir2020energy}, redundancy in the network arises from the number of nodes and the workload on each node. In \cite{zhang2021traffic}, simulation results obtained for network traffic redundancy showed its impact with network size, number of peers, and routing length. A linear relation was found between the total number of peers and the traffic redundancy in the Bitcoin network, with over 98\% of network traffic being redundant showing inefficiency in the current Bitcoin broadcasting algorithm. Every 1000 nodes were shown to increase the effective traffic by 0.3 GB, while the total traffic increased by 24 GB showing a redundancy of 23.7 GB. In addition, the study found a positive correlation between the routing path length and traffic redundancy in the network, demonstrating that denser networks consisting of shorter routing lengths have less redundant traffic.

\subsection{Mining Devices}

In \cite{economist}, the authors argued that if all mining facilities utilized the highly efficient ASIC-based mining devices as done in the KnCMiner Facility in Sweden, the overall Bitcoin mining process would consume nearly 1.46 TWh worldwide which is much lower than the current estimates of 184 TWh \cite{de2021bitcoin}, 135.12 TWh \cite{digi2021bitcoin} and 69.63 TWh \cite{uoc2021bitcoin} for 2021.This discrepancy demonstrates that inefficient mining devices are being used worldwide. Thus, a major contributor of energy consumption is the use of inefficient mining devices \cite{houy2019rational,kufeouglu2019bitcoin}. 

\begin{table}[!t]
\vspace{0.05in}
\centering
\caption{Performance metrics of different mining devices (Sources: \cite{stoll2019carbon,kufeouglu2019bitcoin,economist}).}
\resizebox{\columnwidth}{!}{
\begin{tabular}{|l|l|l|l|} \hline
\textbf{Hardware type} & \textbf{Mining rate (GH/s)} & \textbf{Efficiency (J/GH)} & \textbf{mEC (TWh)} \\ \hline
CPU                    & 0.01                        & 9000     & 11,000                  \\ \hline
GPU                    & 0.2 – 2                     & 1500 – 400     & 3,000               \\ \hline
FPGA                   & 0.1 – 25                    & 100 – 45     & 250                 \\ \hline
ASIC                   & 44,000                      & 0.05     & 1.46                      \\ \hline        
\end{tabular}}
\label{devices}
\end{table}

\begin{figure}[t]
        \vspace{-0.1in}
        \centering
        \includegraphics[width = \columnwidth]{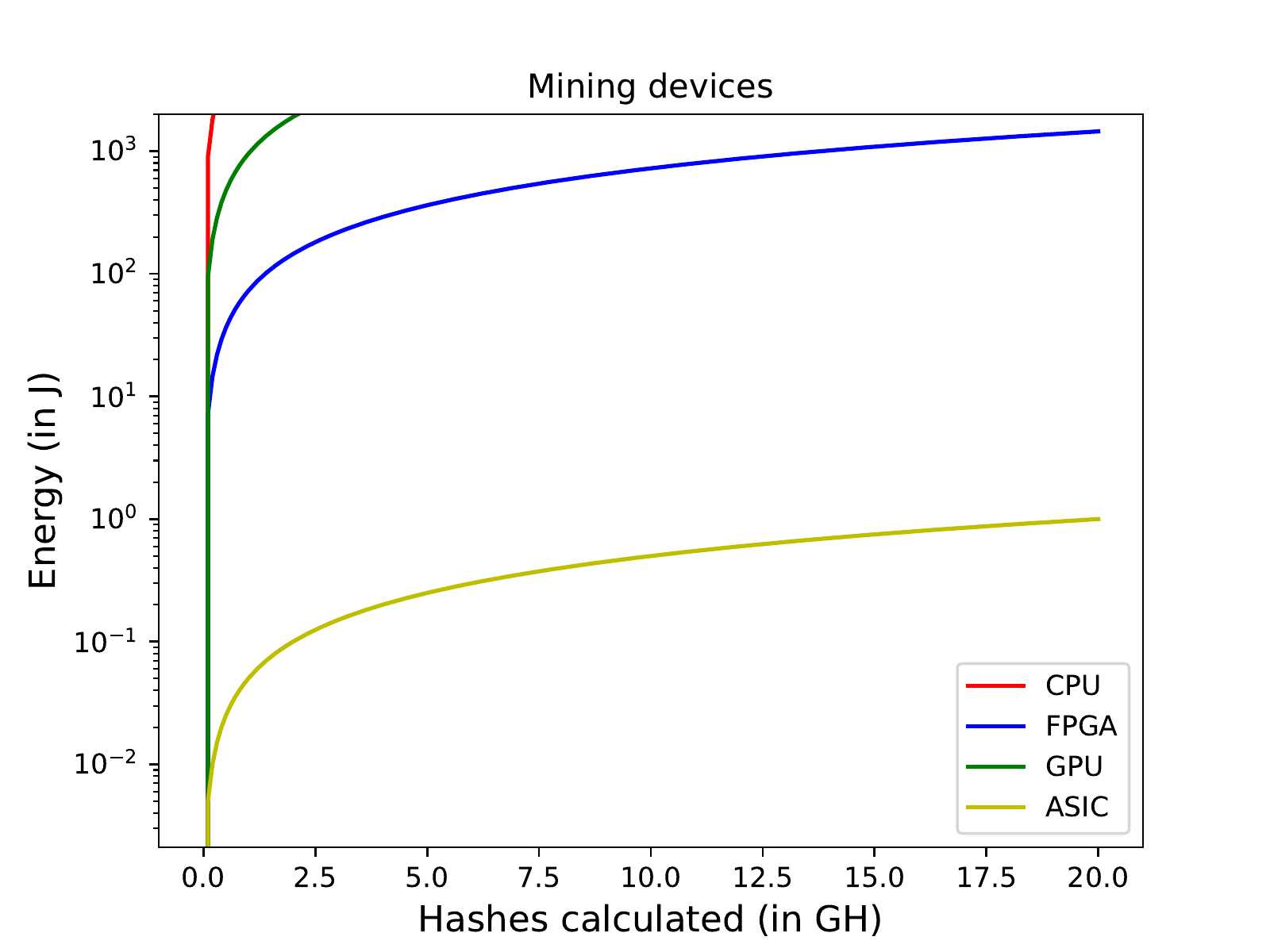}
        \caption{{Logarithmic plot of electrical energy versus hashes calculated for Application-Specific Integrated Circuit (ASIC), Field-Programmable Gate Array (FPGA), Graphics Processing Unit (GPU) and Central Processing Unit (CPU) devices \cite{houy2019rational}.}}
        \label{log-device}
        \vspace{-0.1in}
\end{figure}

\begin{figure*}[!t]
        \vspace{-0.1in}
        \centering
        \includegraphics[width = 2\columnwidth]{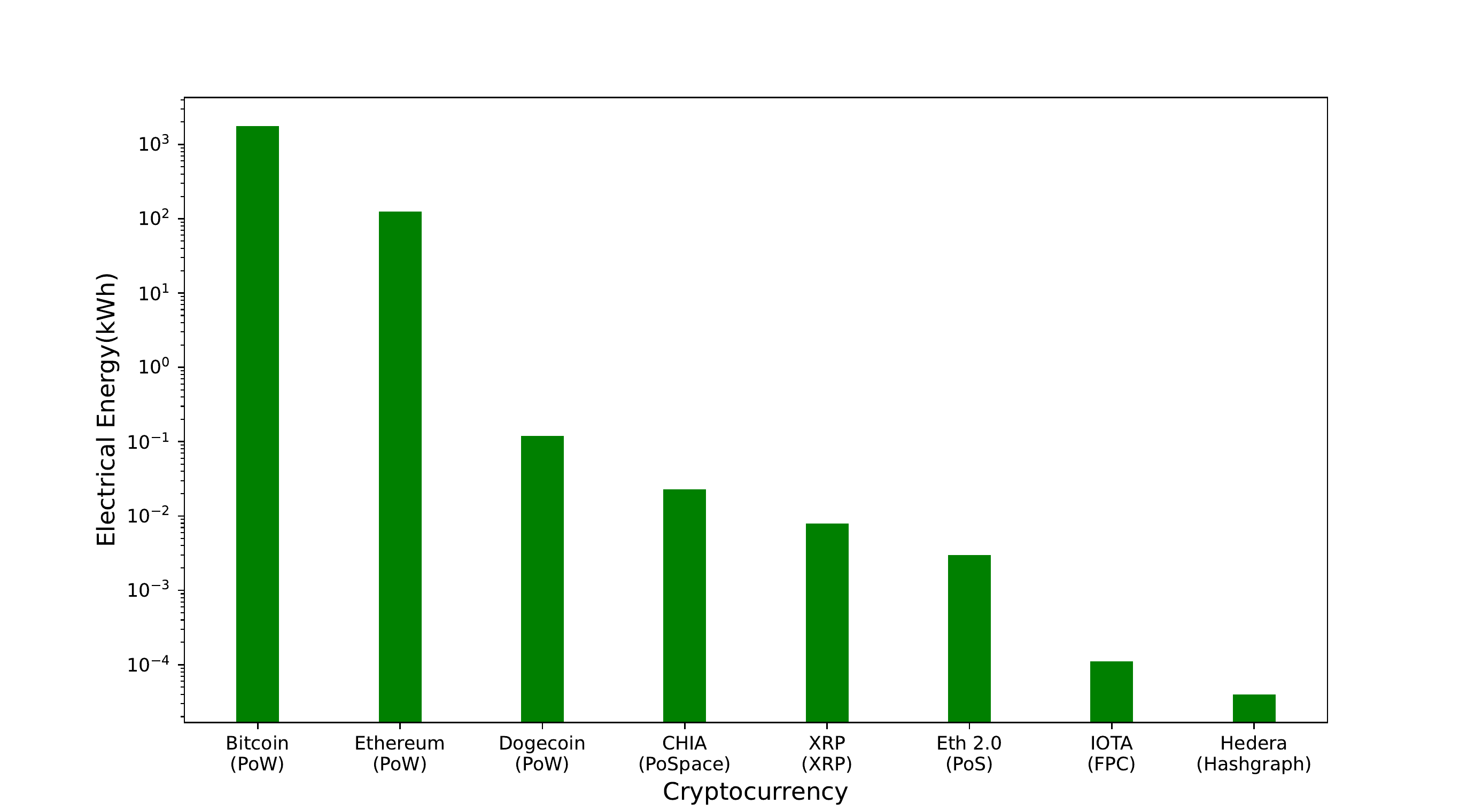}
        \caption{Energy consumption per transaction for various cryptocurrencies and their consensus mechanisms.}
        \label{fig:cryptocomp}
        \vspace{-0.1in}
\end{figure*}

Due to the increasing difficulty of mining, several devices have been used since the introduction of Bitcoin starting from CPU in 2009, to GPU in 2010, FPGA in 2011 and ASIC since 2013 \cite{stoll2019carbon}. Table-\ref{devices} presents these devices along with their hash rates, efficiencies \cite{houy2019rational}, and minimum total energy consumption \cite{kufeouglu2019bitcoin,economist}. The total energy consumption corresponds to the amount of energy used when only that type of device is used for Bitcoin mining worldwide. From the table, we note that CPU provides the least computational power calculated in giga-hashes (GH) at 0.01 GH/s at 9000 J/GH, while consuming the most energy per GH, while ASIC-based devices used in the study provide the highest computational power with 44,000 GH/s at an efficiency of 0.05 J/GH. Figure-\ref{log-device} is a logarithmic-plot of log-energy consumption against the number of GH computed. The plot depicts the high energy consumption of CPU devices followed by GPU, FPGA and ASIC being the most efficient among them. It is important to note that ASIC-based devices provide 40,000 to 200,000 times the computational power of GPUs which can be seen from in Table-\ref{devices}. They create a problem of centralization of computational power \cite{mariem2020all}. ASIC-resistant algorithms remove the benefit of using ASIC-based devices, because reaching a solution for these algorithms with ASIC deices is either impossible or comparable to GPUs. Such algorithms, for instance, X16Rv2 in Ravencoin and Ethash in Ethereum, force miners to use general purpose and cheaper devices such as GPUs, which cause an over-proportionate amount of energy consumption \cite{gallersdorfer2020energy}.

In 2019, the authors of \cite{kufeouglu2019bitcoin} investigated the power demand of Bitcoin mining by considering the performance of 269 mining hardware devices (111 CPU, 111 GPU, 4 FPGA and 43 ASIC) in a 160 GB Bitcoin network. They used data published by the manufacturer in whitepapers corresponding to the device, and also the user-benchmark \cite{userbenchmark} and pass-mark \cite{passmark} websites for manufacturer reliability. The study also considered mining pools \cite{miningpools} to make estimates using the regional electricity costs. Two metrics were defined namely the Minimum Energy Consumption (mEC) and Maximum Energy Consumption (MEC), corresponding to the energy consumption of the most efficient and the least efficient devices respectively relative to context. Calculations of the mEC showed that in comparison to the global energy demand of 23,000 TWh, continued use for only CPU devices alone would consume a minimum of 11,000 TWh of electrical energy. GPU and FPGA devices were shown to consume a minimum of nearly 3,000 TWh, 250 TWh respectively while ASIC devices consumed the least among all devices. The minimum and maximum power demands for all devices were shown to be 2 GW and 6 GW respectively. 

\subsection{Sources of Energy}

The annual carbon footprint of Aluminium mining has been estimated at 90 Mt$CO_2$ \cite{de2021bitcoin} and that of Oman is 68.8 Mt$CO_2$ as Table-\ref{$CO_2$-comp} shows. From our earlier discussion on global comparisons of carbon footprints of cryptocurrencies in Section-\ref{sec:background}, considering the latest emission estimates of upto 90 Mt$CO_2$ \cite{de2021bitcoin} and 64.18 Mt$CO_2$ \cite{digi2021bitcoin} for Bitcoin and 26.13 Mt$CO_2$ for Ethereum \cite{digi2021ethereum} respectively, it is alarming to see nation-level and industry-level carbon emissions from relatively nascent transaction systems. 

The earliest research on the impact of energy consumption on ecological footprints \cite{chen2007ecological} found the negative impacts of fossil fuels on the environment. Subsequently, research indicated a negative impact on the ecological footprint due to the excessive use of fossil fuels by the pulp production industry in the Canadian Prairies \cite{kissinger2007wood}. It is important to note that the results of these studies can also extend to non-renewable energy based cryptocurrency mining. While the sources of energy themselves do not cause the over-proportionate electrical energy consumption in PoW blockchains, the use of non-renewable energy sources leads to high carbon footprints \cite{liu2021selection}. 

Due to the cryptocurrency networks being distributed, it is difficult to obtain an accurate share of renewable and non-renewable sources of energy during mining \cite{koomey2019estimating}. Additionally, there is also uncertainty in estimations based on the mining devices used \cite{houy2019rational}. While some studies \cite{bendiksen2018bitcoin,li2018life} argue that the main source of energy for cryptocurrencies is renewable with a share of nearly 80\%, the 3rd GCBS \cite{blandin20203rd} shows a 61\% reliance on non-renewable sources of energy. Statistics of cryptocurrency mining in China show a 58\% and 42\% split of hydro-energy and coal-heavy power generation respectively according to a recent study \cite{stoll2019carbon}. The research has estimated an adjustment emission factor of 550 g/kWh for China by considering a weighted average of the hydro-rich and coal-heavy provinces of Sichuan and Inner Mongolia. Considering the mining pool share based on hashrate (number of hashes computed per second) of 46\% in China \cite{uoc2021bitcoin} and the prediction of 130 Mt$CO_2$ of the Bitcoin network by 2021 in China alone \cite{jiang2021policy}, along with the continued use of fossil fuels, there is a real threat to the environment all of which make the expected rise of 2\degree C contributed by Bitcoin within the next few decades a real possibility \cite{mora2018bitcoin}. It is therefore necessary to find green solutions to minimize the $CO_2$ emissions of cryptocurrencies.

\section{Solutions}
\label{sec:solutions}

As we have discussed in the previous section, current day cryptocurrencies impose problems of high energy consumption and $CO_2$ emissions because of various reasons such as the PoW, redundancy, device efficiency and sources of energy. This section explores and recommends solutions to address these issues by providing examples from past works in individual areas of alternative consensus mechanisms and redundancy reduction. It also discusses some of the most popular and effective mining devices as of July 2021, and conducts an in depth analysis of renewable energy sources in top mining areas as alternatives to fossil fuels to reduce the carbon footprints of cryptocurrencies.

\begin{table*}[t]
\centering
\caption{Proposed ASIC-based mining devices for cryptocurrency mining as of March 2021 \cite{tech2021mining}.}
\resizebox{\textwidth}{!}{
\begin{tabular}{|l|r|r|r|r|c|}
\hline
\multicolumn{1}{|c|}{\textbf{Device}} & \multicolumn{1}{c|}{\textbf{\begin{tabular}[c]{@{}c@{}}Cost  (\$)\end{tabular}}} & \multicolumn{1}{c|}{\textbf{\begin{tabular}[c]{@{}c@{}}Hashrate  (TH/s)\end{tabular}}} & \multicolumn{1}{c|}{\textbf{\begin{tabular}[c]{@{}c@{}}Power  (W)\end{tabular}}} & \multicolumn{1}{c|}{\textbf{\begin{tabular}[c]{@{}c@{}}Efficiency  (J/TH)\end{tabular}}} & \multicolumn{1}{c|}{\textbf{\begin{tabular}[c]{@{}c@{}} MAC  (GWh)\end{tabular}}} \\ \hline
Whatsminer M32-70                     & 6,200                                                                              & 70                                                                                       & 3,360                                                                              & 48                                                                                             & 29.43                                                                                                 \\ \hline
Antminer S7                           & Variable                                                                           & 4.73                                                                                     & 1,293                                                                              & 273.36                                                                                         & 11.32                                                                                                 \\ \hline
AvalonMiner 1246                      & Variable                                                                           & 90                                                                                       & 3,420                                                                              & 38                                                                                             & 29.95                                                                                                 \\ \hline
WhatsMiner M32-62T                    & 1,075                                                                              & 62                                                                                       & 3,348                                                                              & 54                                                                                             & 29.32                                                                                                 \\ \hline
AvalonMiner A1166 Pro                 & 2,199                                                                              & 81                                                                                       & 3,400                                                                              & 41.97                                                                                          & 29.78                                                                                                 \\ \hline
\end{tabular}}

\label{asic-device}
\end{table*}

\subsection{Alternative Consensus Mechanisms}

Out of the four issues we have discussed above, the consensus mechanism is the biggest contributor to the energy consumption in current cryptocurrencies that use PoW. Thus, one viable option would be to explore other consensus mechanisms which are more energy-efficient than PoW. Figure-\ref{fig:cryptocomp} provides the Electrical Energy Consumption per transaction (EEC/trans) for various cryptocurrencies compiled from the studies \cite{digi2021bitcoin,digi2021ethereum,tgr2021data,platt2021energy}. 

One of the most promising substitutes for the PoW is the Proof of Stake (PoS) consensus mechanism which was first used in Peercoin \cite{kingppcoin} as an energy saving alternative of PoW. It has also been proposed in Ethereum 2.0 which is discussed in Section-\ref{sec:case}. In PoS, the proof is derived from stakes, i.e. contributions of miners to the blockchain, instead of computational power. This removes the computational race involved in the PoW thereby reducing energy consumption and $CO_2$ emissions during mining \cite{nguyen2019proof}. In a study from 2014 \cite{bentov2014proof}, the authors extended the PoW using PoS and proposed the Proof of Activity (PoA) which provided reduced network communication and storage requirements without compromising on security. Proof of Burn (PoB) is another low energy consuming consensus mechanism \cite{karantias2020proof}. Miners reach a consensus by "burning" coins and permanently remove them from circulation. This process is initiated by miners on virtual mining rigs instead of physical mining devices. A miner's mining power increases when the number of coins burned, and not based on computational power. PoB has been proven to be sustainable and highly decentralized, and is implemented in cryptocurrencies such as SlimCoin.

Hedera is an exo-friendly cryptocurrency with a highly efficient consensus mechanism called Hashgraph \cite{baird2016swirlds}, based on the gossip protocol. Participants in the blockchain relay novel information (called gossip), and the collaborative gossip history is stored as a hashgraph, which each member in the network uses to comes to a consensus based on their knowledge of what another node might know. The authors f \cite{popov2021fpc} proposed probabilistic mechanism called the Fast Probabilistic Consensus (FPC). It is used in the cryptocurrency IOTA. It is a highly efficient and secure binary voting protocol wherein a set of nodes can come to consensus on the value of an individual bit instead of consensus through computation. A trust-based mechanism called the XRP Consensus \cite{chase2018analysis} has also been proposed, in which the participants reach an agreement without complete consensus among all members of the network. Hashgraph, FPC and XRP do not require high computational power and therefore consume substantially lower energy than PoW, which can also be seen in Figure-\ref{fig:cryptocomp}. 

The authors of \cite{luu2015scp} proposed the Stellar Consensus Protocol (SCP) based on the Byzantine Agreement \cite{dolev1983authenticated}. It removes time-limitations for the processing of blocks by enabling flexibility in the PoW-difficulty parameters and processes several blocks in parallel. Increased computational power therefore increases the throughput of the system, thereby increasing the scalability and sustainability because there are more blocks processed in the same amount of time, making the energy consumption proportionate to the outcome obtained. SCP is used in the Pi Network \cite{pinet} which we discussed in more detail in Section-\ref{sec:case}. 

\begin{figure*}[!t]
        \vspace{-0.1in}
        \centering
        \includegraphics[width = 2\columnwidth]{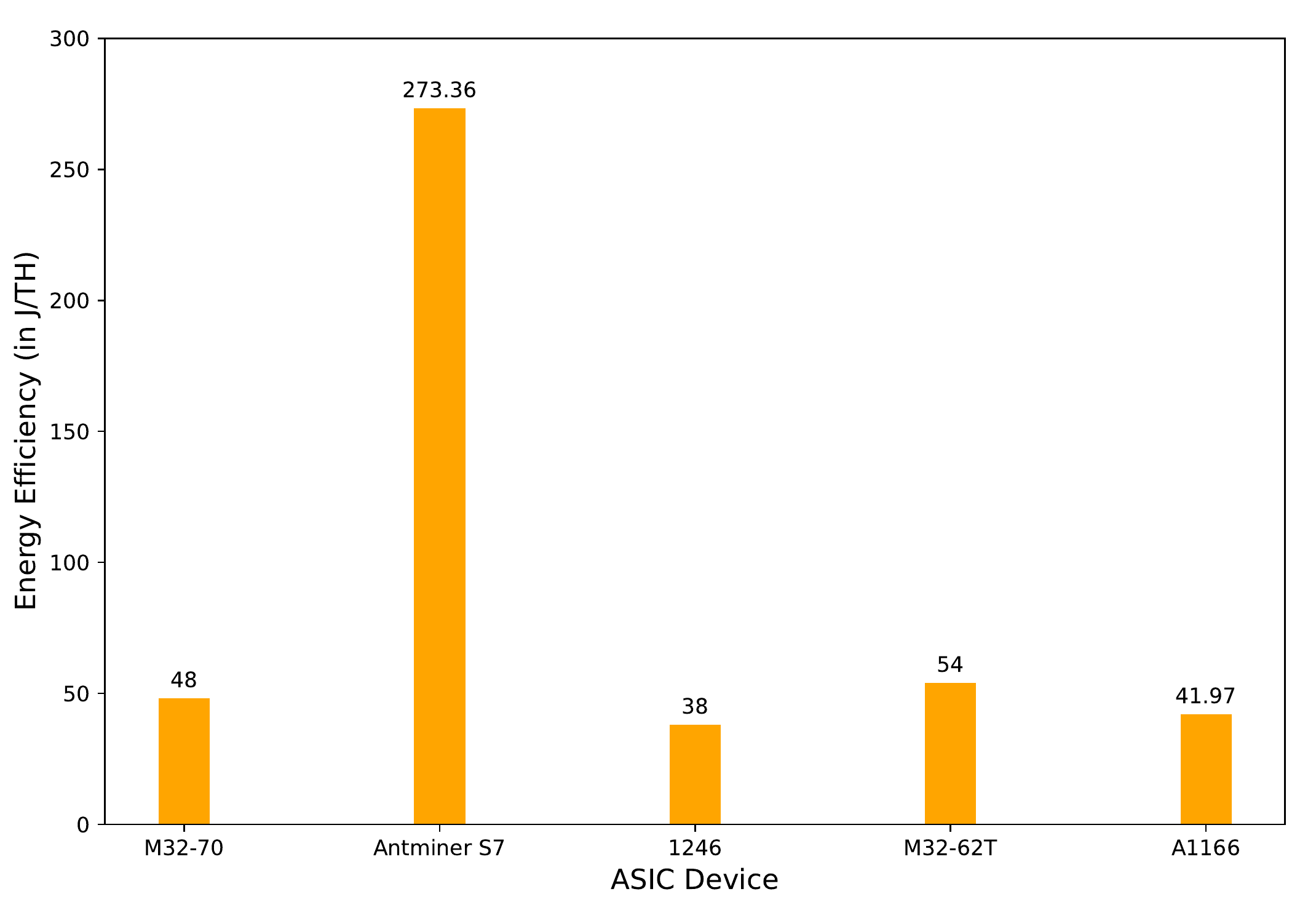}
        \caption{Energy efficiency (energy consumption per TH) of various ASIC devices.}
        \label{asic}
        \vspace{-0.1in}
\end{figure*}

Several storage-based consensus mechanisms have been proposed. The authors of \cite{miller2014permacoin} proposed a consensus mechanism based on distributed storage called the Proof of Retrievability (PoR). However, because the proposed scheme lacks a leader node election method, a similar PoR-based approach was proposed in \cite{moran2019simple} called the Proof of SpaceTime (PoST). PoST proves that useful data was stored for a certain amount of time, and it is thus a storage power-based consensus mechanism. PoST consumes less energy because the difficulty of the proof can be changed by extending the time-period of data stored instead of computational capacity. Another storage share based consensus mechanism called the Proof of Space (PoSpace) was adopted in SpaceCoin \cite{park2015spacecoin} and CHIA \cite{cohen2019chia}. PoSpace requires little computation power and can be run on any free computer with free disk space and an Internet connection.

The authors of \cite{ball2017proofs} recommend the adoption of useful Proofs of Work (uPoW) based on the Orthogonal Vectors (OV) problem. They explain usefulness as the allocation of computational tasks to the miners such that the solutions for the tasks can be reconstructed verifiably and quickly from the miners' response. uPoW converts the amount of wasteful work in PoW into useful work without compromising on hardness. Research on Resource Efficient Mining (REM) \cite{zhang2017rem} for Bitcoin proposed the REM framework using trusted hardware (Intel SGX) and developed the first complete implementation of SGX-blockchain with a computational overhead of 5-15\%. This mechanisms is similar to the uPoW \cite{wang2019survey}. Clients supply their workloads as tasks to the SGX protected enclave. The truthfulness guaranteed feature of the attestation service in SGX verifies and measures the software running in the enclave. The enclave randomly decides which computational task leads to a valid proof for the block.

\subsection{Redundancy Reduction Techniques}

Among the methods proposed in the literature for reducing storage redundancy in blockchain networks, a promising one relies on "sharding", i.e. breaking the network into sub-parts called "shards" based on the consensus mechanism and updating the transactions within the bounds of each shard \cite{sedlmeir2020energy}. In \cite{zamani2018rapidchain}, the authors conducted research on scaling blockchain via sharding, and proposed a stable sharding technique with a low failure rate. The concept of sharding has also been proposed for Ethereum 2.0 which we discuss in Section-\ref{sec:case}. While the division of blockchain networks into shards is difficult because of the decentralization of computational power in the PoW, it can be done based on the proportions of stakes and storage in the case of PoS and PoSpace respectively \cite{nguyen2019proof,wang2019survey}. In \cite{jia2018elasticchain}, the authors proposed another method (called ElasticChain) to reduce redundancy. In ElasticChain, the nodes of the chain store a part of the complete ledger based on a duplicate ratio regulation algorithm. The research shows stability, security and fault tolerance at the same level as the current blockchain design, while improving its storage scalability. The authors of \cite{xue2020semantic} proposed a different approach with Semantic Differential Transaction (SDT) to reduce redundancy in the integration of Building Information Modeling (BIM) and blockchain. SDT captures local changes in an information model as BIM Change Contracts (BCC) at 0.02\% the size of Industry Foundation Classes (IFC) (the standard of ensuring interoperability across BIM platforms, safeguarding them in a blockchain and restoring them when needed). SDT thus reduces redundancy in BIM-blockchain systems. A study on network traffic redundancy \cite{zhang2021traffic} recommends reducing the average routing path lengths between two nodes in order to reduce traffic redundancy in the Bitcoin network.

Another category of methods proposed to reduce operational redundancy in blockchains lies in the use of Zero Knowledge Proofs (ZKP) such as SNARKS \cite{yang2020zero,pinto2020introduction}. ZKP does not require complex encryption. It increases privacy of users by avoiding the disclosure of personal information as is the case in public blockchains such as Bitcoin. Additionally, it provides security while increasing the scalability and throughput of the cryptocurrency network, thereby making it more energy-efficient. The methodology proposed by the authors of \cite{ben2019scalable} uses ZKP to reduce the time needed to prove and verify large sequential computations in comparison to other current ZKP implementations \cite{ishai2015zero}.

\begin{figure*}[t]
        \vspace{-0.1in}
        \centering
        \includegraphics[width = 1.5\columnwidth]{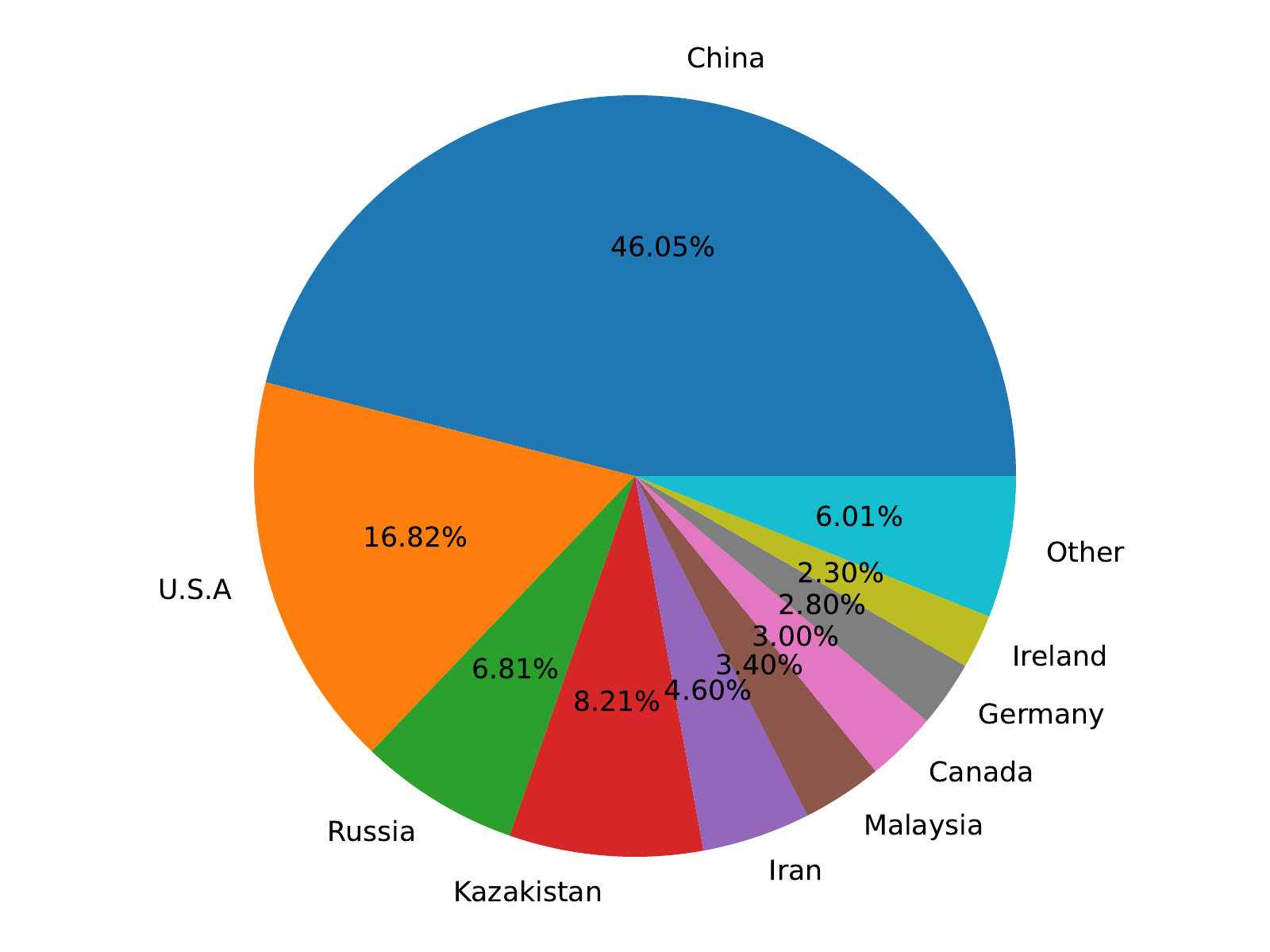}
        \caption{Mining share based on hashrates  \cite{uoc2021bitcoin}.}
        \label{mining-share}
        \vspace{-0.1in}
\end{figure*}

\subsection{Choice of Mining Device}

While efficient devices will help reduce energy costs regardless of the consensus mechanism, if the PoW continues to be in use, it becomes imperative to use highly efficient devices such as ASIC \cite{houy2019rational}. As Table-\ref{devices} and Figure-\ref{log-device} show, ASIC-based devices consume the least amount of energy per hash, and provide the highest computational power with a hash rate of 40,000 GH/s at 0.05 J/GH. Studies have shown that the use of ASIC devices as done in the KnCMiner facility in Boden, Sweden can reduce the worldwide annual energy consumption by mining to 1.46 TWh \cite{economist,kufeouglu2019bitcoin}. 

In \cite{tech2021mining}, the author discusses the top five ASIC-based mining devices as of March 2021 which include Whatsminer M32-70, Antminer S7, AvalonMiner 1246, WhatsMiner M32-62T, and AvalonMiner A1166 Pro. Table-\ref{asic-device} presents the cost, hashrate, power consumption, efficiency and Maximum Annual Consumption (MAC) for each of these devices. Figure-\ref{asic} shows that among the most popular available options, Antminer S7 is the least efficient device, with the energy consumption per terra-hash (TH) of 273.36 J/TH. The other four devices have comparable efficiencies, with AvlonMiner 1246 being the most energy-efficient at 38 J/TH. In addition to the efficiencies, we have also calculated the MAC (in GWh) of these devices as follows:

\begin{equation}
 MAC = \frac{P\times24\times365}{10^6}
\end{equation}

wherein, $P$ (in W) is the power consumption of the device which is multiplied with the total number of hours in a year as $24\times365$ to obtain the annual energy consumption equivalent. The table shows that Antminer S7 consumed the least amount of energy, while providing the least efficiency among the five options. The other four ASIC-based mining devices have comparable MAC ranging from 29.32 GWh to 29.95 GWh, thereby further demonstratingthat the best choice is AvlonMiner 1247 based on its efficiency.

\begin{table*}[!t]
\centering
\caption{Mining shares, renewable energy capacities installed and evaluation metrics: Estimated Energy Consumptions (EEC), Maximum Energy Generation (MEG), Renewable Capacity Ratio (RCR) for major Bitcoin mining regions.}
\resizebox{\textwidth}{!}{
\begin{tabular}{|l|r|r|r|r|r|r|r|r|r|r|r|r|}
\hline
\multicolumn{1}{|c|}{\multirow{2}{*}{\textbf{Region}}} & \multicolumn{2}{c|}{\textbf{Bitcoin mining \cite{uoc2021bitcoin,digi2021bitcoin}}}                                          & \multicolumn{7}{c|}{\textbf{Renewable energy capacity installed (MW \cite{irena2020renewable}}}                                                                                                                                                                                                                  & \multicolumn{1}{c|}{\multirow{2}{*}{\textbf{\begin{tabular}[c]{@{}c@{}} MEG \\ (TWh)\end{tabular}}}} & \multicolumn{1}{c|}{\multirow{2}{*}{\textbf{RCR}}} & \multicolumn{1}{c|}{\multirow{2}{*}{\textbf{Relative RCR}}} \\ \cline{2-10}
\multicolumn{1}{|c|}{}                                 & \multicolumn{1}{c|}{\textbf{Share (\%)}} & \multicolumn{1}{c|}{\textbf{EEC (TWh)}} & \multicolumn{1}{c|}{\textbf{Total capacity}} & \multicolumn{1}{c|}{\textbf{Hydropower}} & \multicolumn{1}{c|}{\textbf{Wind}} & \multicolumn{1}{c|}{\textbf{Solar}} & \multicolumn{1}{c|}{\textbf{Bioenergy}} & \multicolumn{1}{c|}{\textbf{Geothermal}} & \multicolumn{1}{c|}{\textbf{Marine}} & \multicolumn{1}{c|}{}                                         & \multicolumn{1}{c|}{}                                & \multicolumn{1}{c|}{}                                         \\ \hline
World                                                  & 100                                      & 135.12                                     & 2,799,094                                    & 1,331,889                                & 733,267                            & 713,970                             & 126,557                                 & 14,050                                   & 527                                  & 24520.06                                                      & -                                                    & -                                                             \\ \hline
China                                                  & 46                                       & 62.15                                      & 894,879                                      & 370,160                                  & 281,993                            & 254,335                             & 18,687                                  & 0                                        & 5                                    & 7839.14                                                       & 126.12                                               & 0.4134                                                        \\ \hline
U.S.A                                                  & 16.8                                     & 22.70                                      & 292,065                                      & 103,058.00                               & 117,744                            & 75,572                              & 12,372                                  & 2,587                                    & 0                                    & 2558.48                                                       & 112.70                                               & 0.3694                                                        \\ \hline
Kazakhstan                                             & 8.2                                      & 11.07                                      & 4,997                                        & 2,785                                    & 486                                & 1,719                               & 8                                       & -                                        & -                                    & 43.77                                                         & 3.95                                                 & 0.0129                                                        \\ \hline
Russia                                                 & 6.8                                      & 9.18                                       & 54,274                                       & 51,811                                   & 945                                & 1,428                               & 1,370                                   & 74                                       & 2                                    & 475.44                                                        & 51.74                                                & 0.1696                                                        \\ \hline

Iran                                                   & 4.6                                      & 6.21                                       & 12,922                                       & 13,233                                   & 303                                & 414                                 & 12                                      & -                                        & -                                    & 113.19                                                        & 18.21                                                & 0.0597                                                        \\ \hline
Malaysia                                               & 3.4                                      & 4.59                                       & 8,699                                        & 6,275                                    &                                    & 1,493                               & 931                                     & -                                        & -                                    & 76.20                                                         & 16.58                                                & 0.0543                                                        \\ \hline
Canada                                                 & 3                                        & 4.05                                       & 101,188                                      & 81,058                                   & 13,577                             & 3,325                               & 3,383                                   & -                                        & 20                                   & 886.40                                                        & 218.67                                               & 0.7168                                                        \\ \hline
Germany                                                & 2.8                                      & 3.78                                       & 131,739                                      & 10,720                                   & 62,184                             & 53,783                              & 10,364                                  & 40                                       & -                                    & 1154.03                                                       & 305.02                                               & 1.0000                                                        \\ \hline
Ireland                                                & 2.3                                      & 3.10                                       & 4,685                                        & 529                                      & 4,300                              & 40                                  & 107                                     & -                                        & -                                    & 41.04                                                         & 13.20                                                & 0.0432                                                        \\ \hline
Other                                                  & 6                                        & 8.24                                       & 1,293,646                                    & 692,260                                  & 251,735                            & 321,861                             & 79,323                                  & 11,349                                   & 500                                  & 11332.33                                                      & -                                                    & -                                                             \\ \hline
\end{tabular}}
\label{renewable}
\end{table*}

\subsection{Renewable Sources of Energy}

Considering the high electrical energy consumption in current PoW blockchains, we need to address the impact of their emissions and the deterioration of the ecological footprint. Research shows a reduction in $CO_2$ emission by using renewable sources of energy \cite{dong2020renewable}. Sustainable Development Goals (SDG) for economic growth and trade provided by a study on renewable and non-renewable energy and their impact \cite{destek2020renewable} recommends the transition from fossil fuels to renewable energy sources, implementation of environmental friendly production processes, enforcement of green trade, education, and creating awareness. While these recommendations have been provided for sustainable economic growth and trade in general, they are also applicable to cryptocurrencies. In \cite{liu2021selection}, the authors show that legal criteria, and the continuity and cost of electrical energy supply are the most important factors considered to decide the location of cryptocurrency mining operations. The study concluded that wind and solar energy are the best energy alternatives for blockchain networks. The use of these renewable energy sources will make the high energy consumption in PoW cryptocurrencies more environmental friendly. Subsequently it is highly recommended that countries with high cryptocurrency mining activity should invest in the use of renewable energy.

Figure-\ref{mining-share} shows the distribution of mining shares based on hashrates as of July 2021. We obtain the data from the Cambridge Bitcoin Energy Consumption Index \cite{uoc2021bitcoin}. The major Bitcoin mining countries of the world are China (46\%), U.S.A (16.8\%), Kazakhstan (8.2\%), Russia (6.8\%), Iran (4.6\%), Malaysia (3.4\%), Canada (3\%), Germany (2.8\%) and Ireland (2.3\%). Considering the Digiconomist \cite{digi2021bitcoin} estimate of 135.12 TWh for Bitcoin, and energy shares of these to be equal to the hashrate shares, we can calculate their Estimated Energy Consumption ($EEC$) as follows:

\begin{equation}
    EEC = \frac{135.12\times Share (\%)}{100} 
\end{equation}

From Table-\ref{renewable}, we note that China alone consumed 62.15 TWh of electrical energy, which is comparable to the electrical energy consumptions of Switzerland (56.35 TWh). It is therefore important for these major mining regions to focus on measures to minimize the environmental degradation and global warming caused by the PoW mining processes. 

\begin{figure*}[!t]
        \vspace{-0.1in}
        \centering
        \includegraphics[width = 2\columnwidth]{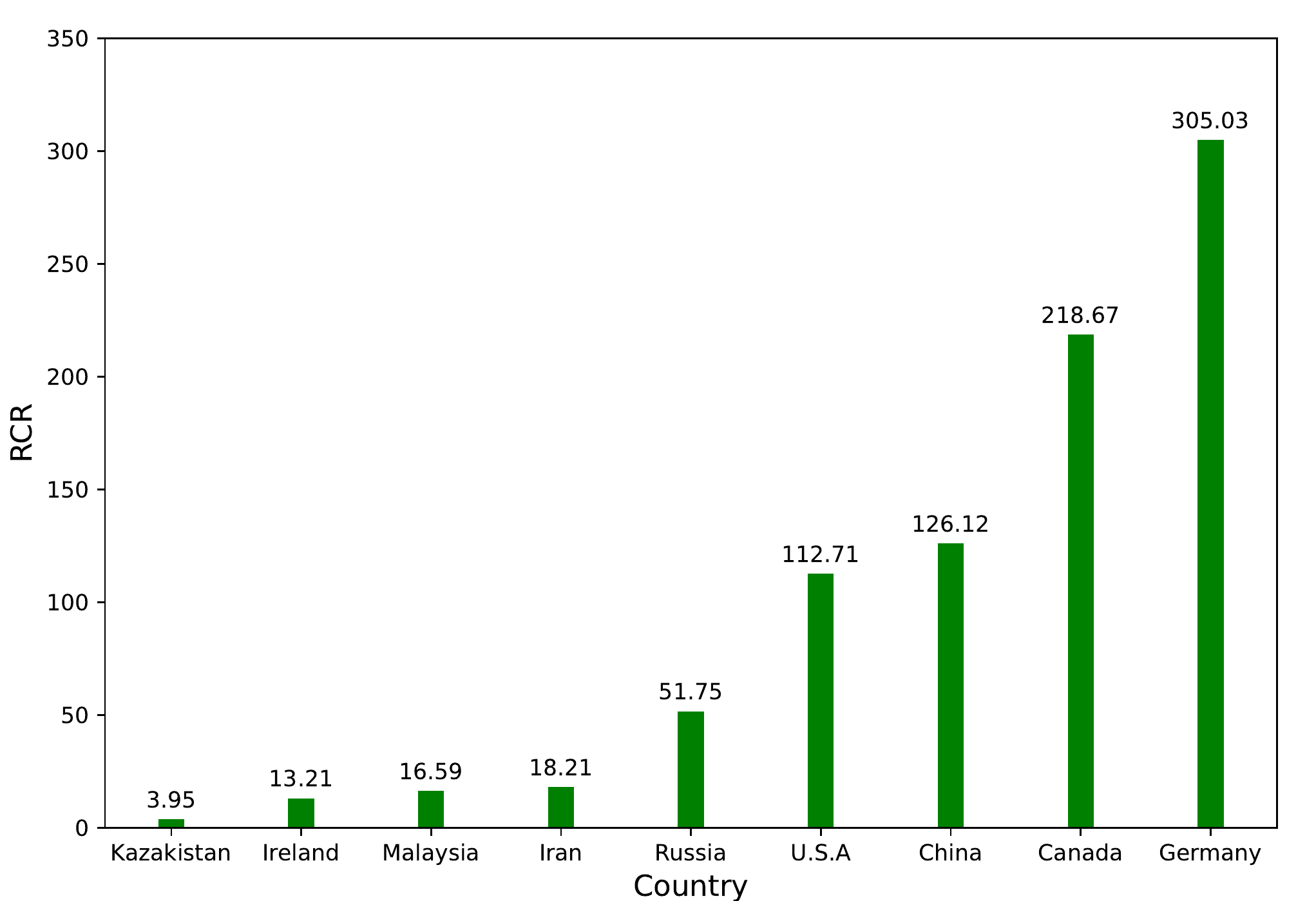}
        \caption{Renewable Capacity Ratio for major Bitcoin mining regions.}
        \label{ratio}
        \vspace{-0.1in}
\end{figure*}

Table-\ref{renewable} presents the data, provided by the International Renewable Energy Agency (IREA) \cite{irena2020renewable}, on various infrastructures based on renewable energy sources as of 2020. The table also presents the Maximum energy Generaction ($MEG$) from renewable sources based on installed renewable capacities for each country. The $MEG$ (in TWh) is calculated using the following equation:

\begin{equation}
    MEG = \frac{Total Capacity\times24\times365}{10^6}
\end{equation}

It is worth noting that the $MEG$ is calculated as the highest possible energy generation per annum using the installed capacity. Additionally the Renewable Capacity Ratio ($RCR$) for each country is calculated as the ratio of the MEG to the EEC following the equation below:

\begin{equation}
    RCR = \frac{MEG}{EEC}
\end{equation}

The $RCR$ provides a proportion of renewable energy available per TWh of energy consumption in Bitcoin mining. High $RCR$ values indicate a higher capacity to allocate renewable energy toward the mining process. From Figure-\ref{ratio}, we deduce that countries such as Germany, Canada, China, and U.S.A have high renewable energy capacities relative to their mining energy consumption in comparison with those such as Kazakhstan, Ireland, Malaysia, Iran, and Russia which do not, making it imperative for these countries to further invest in renewable energy.  

\section{Case Studies}
\label{sec:case}

Sections-\ref{sec:problems} and \ref{sec:solutions} have discussed implementation factors that cause high energy consumption and carbon footprints of cryptocurrencies, and proposed possible solutions respectively. In this section, we explore a few cryptocurrency networks that aim to solve some of the practical limitations of cryptocurrencies such as Bitcoin and Ethereum. As we have discussed earlier, several alternative consensus mechanisms such as the PoS, and redundancy reduction techniques such as sharding, have been proposed to reduce the energy consumption of cryptocurrencies. In this section, we discuss how some recently developed cryptocurrency networks such as Ethereum 2.0 and the Pi Network have adapted these solutions to solve the cryptocurrency energy consumption and carbon footprint problems in the real world. These case studies will provide more insight into the ongoing active research and development, and will shed light on future research directions in this area.

\subsection{Ethereum 2.0}

We briefly described Ethereum in Section-\ref{sec:background}. Several alternate cryptocurrencies have been introduced over time, but none of them have gained as much traction as Bitcoin, with the exception of the PoW cryptocurrency, Ethereum. However, since it also suffers from energy and scalability issues, Ethereum has come up with a major upgrade, called Ethereum 2.0 \cite{eth2.0}. This version aims to resolve issues related to sustainability, scalability, and security. The security aspects are beyond the scope of this paper, hence we focus our discussion on sustainability and scalability:

i. \textbf{Sustainability}: Ethereum 2.0 attempts the energy problem by shifting from the PoW consensus mechanism to the PoS. PoS consumes significantly lower amount of energy because it involves much fewer mathematical calculations and hence has lesser computational requirements. It also provides security against attacks like 51\% attack, and prevents over-centralization of miners as ownership of coins is considered as opposed to share of computational power for reward payouts. This change in consensus algorithm is expected to consume less than 99\% of the current consumption of the PoW algorithm. \cite{ieee2019ethereum}. 

ii. \textbf{Scalability}: The current version of Ethereum is not very scalable due to the increase in network congestion and data redundancy with the addition of nodes and transactions. This increases the energy consumption of the cryptocurrency network in addition to slowing down the speed of the transaction process. With Ethereum 2.0, Ethereum plans to introduce the "Beacon Chain" which implements the concept of sharding. Sharding is a concept where the load on a network is distributed amongst nodes or groups of nodes to reduce network congestion and increase throughput. The release will also include the introduction of 64 new chains, with each chain consisting of a fraction of the nodes validating the transactions. Hence more transactions can be processed in parallel, with the requirement to share the transaction details with only a fraction of the nodes. This reduces redundancy, congestion and energy consumption.

\subsection{Pi Network}
In \cite{pinet}, the authors present an introduction to the Pi Network which addresses the two issues that the Bitcoin network suffers from namely, high energy consumption and centralization of miners.

i. \textbf{Energy efficiency}: The Pi Network uses a modified version of the Stellar Consensus Protocol (SCP) \cite{scprotocol} instead of the highly energy intensive PoW consensus mechanism. While such networks need multiple exchanges among the nodes to reach consensus and can lead to network congestion, they have significantly lower energy requirements.

ii. \textbf{Decentralization}: While the original goal of Bitcoin was to provide a decentralized transaction method, the increase in price and better payoffs has made the network extremely centralized to the extent that around 87\% of the BTCs are owned by 1\% of the nodes. The Pi network allows any user with a mobile phone to mine coins without any need for expensive ASIC devices. Hence it makes mining inexpensive and more widely accessible.

\section{Conclusion}
\label{sec:conc}
This review has shown the alarmingly high electrical energy consumption and carbon footprints of PoW cryptocurrencies such as Bitcoin and Ethereum. When compared the energy consumption of countries around the world, we found that Bitcoin and Ethereum consumed nearly as much energy as countries such as Sweden and Romania respectively. We also found that their $CO_2$ emissions were close to those of Greece and Tunisia respectively. Our analysis of centralized transaction methods has revealed that Visa is much more energy-efficient and has a lower carbon footprint per transaction compared to the cryptocurrencies discussed in this review. The review identified four underlying issues causing these problems, namely, the PoW consensus mechanism, network redundancy, mining devices and sources of energy. We found that, among other possible solutions such as PoSpace, PoST, PoA, uPoW and REM, PoS proves to be the most promising alternative to PoW. We discussed redundancy reduction methods and popular ASIC devices for efficient mining. We compiled a list of popular mining devices available on the market that would be useful to various stakeholders working in the cryptocurrency area. We calculated the maximum possible energy consumption using $MAC$. Additionally, we presented renewable energy capacities for major Bitcoin mining areas, and the defined $RCR$ has showed that it would be easier for major mining countries such as China, U.S.A, Germany and Canada to allocate renewable energy compared to countries such as Russia, Iran, Malaysia, Ireland and Kazakhstan. Finally, we presented two case studies on Ethereum 2.0 and the Pi-Network which plan to use consensus algorithms such as $PoS$ and $SCP$, and concepts such as sharding to distribute the load and reduce redundancy in the cryptocurrency network to reduce the overall energy consumption and carbon footprints. While these networks are still under development, they demonstrate that considerable efforts are being made in this direction to address the real world energy consumption and $CO_2$ issues associated with cryptocurrencies to make them more sustainable and widely acceptable.

\section{Acknowledgment}
This work was also supported by the SERB ASEAN prject CRD/2020/000369 received by Dr. Vinay Chamola. Sherali Zeadally was supported by a 2021-2022 Fulbright U.S. scholar grant award administered by the U.S. Department of State Bureau of Educational and Cultural Affairs, and through its cooperating agency the Institute of International Education (“IIE”).  Further, we thank the anonymous reviewers for their valuable comments which helped us improve the quality and presentation of this work. 

\bibliographystyle{elsarticle-num}
\bibliography{Doc}
~~~\\
~~~\\







\end{document}